\documentclass[aps,prd,twocolumn,nofootinbib]{revtex4-2}

\usepackage{booktabs,makecell,multirow}

\usepackage{graphicx}
\usepackage{float}
\usepackage{color}
\usepackage{ulem}
\usepackage{amsmath}

\usepackage[colorlinks=true, linkcolor=blue, citecolor=blue, urlcolor=blue]{hyperref}

\newcommand*{\RMN}[1]{\uppercase\expandafter{\romannumeral#1}}

\begin{document}

\title{NNLO QCD Corrections to $D$-Wave Spin-Singlet Heavy Quarkonia Decay $\eta_{Q2}\to\gamma\gamma$ via the Principle of Maximum Conformality}

\author{Hua Zhou$^1$}
\email{zhouhua@swust.edu.cn}
\author{Xing-Gang Wu$^2$}
\email{wuxg@cqu.edu.cn}
\author{Qing Yu$^1$}
\email{yuq@swust.edu.cn}
\author{Sheng-Quan Wang$^3$}
\email{sqwang@al.cqu.edu.cn}
\affiliation{$^1$ School of Mathematics and Physics, Southwest University of Science and Technology, Mianyang 621010, P.R. China\\
$^2$ Department of Physics, Chongqing Key Laboratory for Strongly Coupled Physics, Chongqing University, Chongqing 401331, P.R. China\\ 
$^3$ Department of Physics, Guizhou Minzu University, Guiyang 550025, P.R. China}

\date{\today}


\begin{abstract}

In this paper, we perform a comprehensive study of the decay process $\eta_{Q2}\to\gamma\gamma$ for $D$-wave spin-singlet heavy quarkonia up to next-to-next-to-leading-order (NNLO) QCD corrections within the nonrelativistic QCD effective theory. Following its factorization formalism, the total decay width is decomposed into perturbatively calculable short-distance coefficients (SDCs) and nonperturbative $D$-wave long-distance matrix elements (LDMEs). The original NNLO series of SDCs suffers from sizable renormalization and factorization scale uncertainties. To eliminate such inherent scale ambiguities, we adopt the Principle of Maximum Conformality (PMC). We show that recursively applying the renormalization group equations for the running of $\alpha_s$ and $D$-wave LDMEs within the PMC framework yields an effective strong coupling $\alpha_s(Q_\ast)$ consistent with the expansion coefficients, resulting in a scale-invariant perturbative series. The determined PMC scales are $Q_\ast=1.483$ GeV for $\eta_{c2}$ and $Q_\ast=4.246$ GeV for $\eta_{b2}$. By removing divergent renormalon contributions, the PMC naturally improves the convergence of the perturbative series for SDCs. Our PMC predictions for the total decay widths are $\Gamma_{\eta_{c2}\to\gamma\gamma}^{\rm PMC} = 3.322^{+0.899}_{-0.828}\ \text{eV}$ and $\Gamma_{\eta_{b2}\to\gamma\gamma}^{\rm PMC} = 0.0188^{+0.0014}_{-0.0013}\ \text{eV}$. The uncertainties arise from variations of the charm and bottom quark masses $\Delta m_c=\pm 0.07$ GeV, $\Delta m_b=\pm 0.06$ GeV, as well as systematic errors from uncalculated higher-order corrections. The corresponding branching ratios are $\text{Br}(\eta_{c2}\to\gamma\gamma) = \big(7.463^{+2.020}_{-1.860}\big)\times 10^{-6}$ and $\text{Br}(\eta_{b2}\to\gamma\gamma) = \big(6.460^{+0.481}_{-0.447}\big)\times 10^{-7}$.

\end{abstract}

\maketitle

\section{Introduction}

The study of heavy quarkonium systems is crucial for probing both perturbative and nonperturbative aspects of Quantum Chromodynamics (QCD). In the charmonium and bottomonium families, the $\eta_{Q2}$ ($Q=c,b$) state is the only low-lying $D$-wave spin-singlet state that remains experimentally unobserved. The production mechanism of such states is complex, governed by the interplay between the perturbative hard-scattering processes and the nonperturbative hadronization effects. High-precision theoretical predictions for the $\eta_{Q2}$ state are therefore indispensable: they not only deepen our understanding of the fundamental properties of strong interactions, but also provide critical guidance for experimental searches by determining the optimal energy ranges for signal identification and detection.

Experimentally, the $S$-wave spin-singlet quarkonium $\eta_Q$ ($Q=c,b$) states are well established with clear experimental signatures: the $\eta_c(1S/2S)$ and $\eta_b(1S/2S)$ states have been precisely measured with high statistical significance by the BESIII, Belle, and BaBar collaborations. In contrast, experimental searches for the $D$-wave spin-singlet charmonium $\eta_{c2}(1D)$ state are still ongoing. The Belle experiment at KEKB investigated the process $e^+e^-\to\gamma\eta_{c2}(1D)$ at center-of-mass energies of $\sqrt{s}=10.52$, $10.58$, and $10.867$ GeV~\cite{Belle:2021cjk}, while the BESIII experiment searched for $\phi\eta_{c2}(1D)$ production via the channels $e^+e^-\to\phi\chi_{c0},\,\phi\eta_{c2}(1D)$ in the energy range $4.47$–$4.95$ GeV~\cite{BESIII:2024itg}. To date, no significant $\eta_{c2}(1D)$ excess has been observed in either experiment.

Theoretically, the $\eta_{Q2}$ state is predicted to lie close to the open-flavor threshold~\cite{Godfrey:1985xj, Ebert:2002pp, Eichten:2004uh, Barnes:2005pb, Burns:2010qq, Kalashnikova:2010hv}, and the decay $\eta_{c2}\to D\bar{D}$ is forbidden by parity conservation. Accordingly, $\eta_{c2}$ is expected to be a narrow resonance, whose dominant decay modes include electric E1 transitions and hadronic decays~\cite{Ackleh:1991dy, Fan:2009cj}. Earlier theoretical studies~\cite{Fan:2009cj, Eichten:2002qv} have examined the hadronic transition $\eta_{c2}\to\pi\pi\eta_c$ and other decay channels of $\eta_{c2}$. For production mechanisms, $B$-meson decays offer a promising platform to search for charmonium states~\cite{Belle:2001oxr, CLEO:2000moj, Belle:2002bnx, BaBar:2001kpq, BaBar:2002tlb, Belle:2019avj}. In particular, the exclusive decay $B^-\to\eta_{c2}K^-$ has been analyzed within the heavy-quark scattering formalism~\cite{Xu:2016kbn}. Moreover, electromagnetic decays of $\eta_{Q2}$ have been explored via the Bethe-Salpeter method~\cite{Wang:2013nya, Du:2024gna}. The diphoton decay channel $\eta_{Q2}\to\gamma\gamma$ is of great experimental interest, as it yields a clean signature for the identification of $\eta_{Q2}$. 

Currently, the QCD corrections to the total decay width of $\eta_{Q2}\to\gamma\gamma$ have been calculated up to $\mathcal{O}(\alpha_s^2)$-order level within the nonrelativistic QCD (NRQCD) factorization formalism~\cite{Bodwin:1994jh}. After integrating over the phase space, the total decay width for $\eta_{Q2}\to\gamma\gamma$ can be written as~\cite{Yang:2020pyh}:
\begin{eqnarray}
\Gamma_{\eta_{Q2}\to\gamma\gamma} = \frac{1}{2!}\frac{1}{2J+1}\frac{1}{8\pi}
\left(2\left|A^{\eta_{Q2}}_{1,1}\right|^2 + 2\left|A^{\eta_{Q2}}_{1,-1}\right|^2\right),
\end{eqnarray}
where the symmetry relation among different helicity amplitudes, $A_{\lambda_1, \lambda_2} = (-1)^J A_{\lambda_2, \lambda_1}$, is implicitly adopted, and the factor $1/2!$ arises from the indistinguishability of the two final-state photons and the total angular momentum of $\eta_{Q2}$ stat is $J=2$. Here, $A^{\eta_{Q2}}_{\lambda_1,\lambda_2}$ stands for the helicity amplitude of the process $\eta_{Q2}\to\gamma(\lambda_1)\gamma(\lambda_2)$, with photon helicities $\lambda_{1,2}=\pm1$. Up to $\mathcal{O}(v^2)$-order level, the helicity amplitude can be factorized into a nonperturbative long-distance matrix element (LDME) and a perturbatively computable short-distance coefficient, e.g.
\begin{eqnarray}
A^{\eta_{Q2}}_{\lambda_1,\lambda_2} &=& {\cal C}^{\eta_{Q2}}_{\lambda_1,\lambda_2}(\mu_r,\mu_f)
\frac{\langle 0|\chi^\dag{\cal K}_{^{1}D_{2}}\psi(\mu_f)|\eta_{Q2}\rangle}{m_Q^{5/2}},
\label{eq:helicity_amplitude}
\end{eqnarray}
where $m_Q$ is the heavy-quark mass~\footnote{We utilize the scale-invariant pole mass for subsequent discussions such that the RGE-related non-conformal $\{\beta_{i}\}$-terms pertain only to $\alpha_{s}$ or the LDME.}, $v$ is the heavy quark velocity in the quarkonium rest frame, $\mu_r$ and $\mu_f$ denote the renormalization scale and factorization scale, respectively. The short-distance coefficient ${\cal C}^{\eta_{Q2}}_{\lambda_1,\lambda_2}(\mu_r,\mu_f)$ has been computed up to next-to-next-to-leading order (NNLO)~\cite{Yang:2020pyh}. Due to parity conversation, we also have $A_{\lambda_1,\lambda_2}=-A_{-\lambda_1,-\lambda_2}$~\cite{Haber:1994pe}, which leads to $A_{1,-1}=0$. The total decay width then becomes
\begin{eqnarray}
\Gamma_{\eta_{Q2}\to\gamma\gamma} &=&
|{\cal C}^{\eta_{Q2}}_{1,1}(\mu_r,\mu_f)|^2 \frac{|\langle 0|\chi^\dag{\cal K}_{^1D_2} \psi(\mu_f)|\eta_{Q2}\rangle|^2}{40\pi m_Q^5}.
\label{eq4}
\end{eqnarray}
It indicates that the total decay width involves both the renormalization scale $\mu_r$ and the factorization scale $\mu_f$. Following the standard renormalization group invariance, a physical observable should not depend on such input parameters. However, this requirement cannot be naturally satisfied for a fixed-order expansion due to the mismatch between scale-dependent quantities, such as $\alpha_s$ and LDME, and their perturbative coefficients. Therefore, a proper scale-setting method must be introduced. It is noted that the Principle of Maximum Conformality (PMC)~\cite{Brodsky:2011ig, Brodsky:2011ta, Mojaza:2012mf, Brodsky:2012rj, Brodsky:2013vpa} well satisfies this physical requirement~\cite{Wu:2013ei, Wu:2014iba, Wu:2019mky, DiGiustino:2023jiq}. By systematically employing the renormalization group equations (RGEs) to determine the effective values of scale-running parameters from the original perturbative series, the resulting PMC series becomes conformal series and independent of renormalization schemes and scales. In this work, we adopt the PMC single-scale method~\cite{Shen:2017pdu} to analyze the NNLO perturbative series for $\Gamma(\eta_{Q2}\to\gamma\gamma)$. This treatment also provides a solid and accurate foundation for estimating uncalculated higher-order contributions~\cite{Du:2018dma, Shen:2022nyr}.

\section{Eliminating Factorization and Renormalization Scale Dependence}

To eliminate the dependence on $\mu_f$, we need to perform the matching between the short-distance coefficients $\mathcal{C}^{\eta_{Q2}}_{1,1}(\mu_r,\mu_f)$ and the LDME $\langle 0|\chi^\dagger \mathcal{K}_{^1D_2} \psi(\mu_f)|\eta_{Q2}\rangle$. Specifically, the $\mu_f$ evolution of the LDME is obtained by evaluating the nonrelativistic heavy-quark pair production/annihilation current, which for the $^1D_2$ state is defined as~\cite{Luke:1999kz, Hoang:2006ty, Manohar:1999xd, Manohar:2000hj}
\begin{displaymath}
(j^{S=0}_{J=2, L=2})^{ij}
= \psi_{\mathbf{p}}^\dagger \left( p^i p^j - \frac{\mathbf{p}^2}{n}\delta^{ij} \right) (i\sigma_2) \chi^*_{-\mathbf{p}}.
\end{displaymath}
Here, $\psi_{\mathbf{p}}$ and $\chi_{\mathbf{p}}$ denote the Pauli spinor fields of fermions with momentum $\mathbf{p}$, and $\sigma$ stands for the Pauli matrices.

Heavy quarkonium systems feature a distinct scale hierarchy $m_Q \gg m_Q v \gg m_Q v^2$. The velocity nonrelativistic QCD (vNRQCD) effective theory~\cite{Luke:1999kz} is constructed to describe such multiscale hierarchies. To this end, the RGE for the LDME is formulated in terms of the dimensionless subtraction velocity $v$, enabling a unified treatment of the soft scale $m_Q v$ and the ultrasoft scale $m_Q v^2$ in nonrelativistic systems. The $\mu_f$ evolution is then derived by performing matching for matrix elements at the hard scale $v=1$. The annihilation current entering the matrix element $\langle 0|\chi^\dagger \mathcal{K}_{^1D_2}\psi(\mu_f)|\eta_{Q2}\rangle$ is the Hermitian conjugate of the aforementioned vNRQCD production current $(j^{S=0}_{2,2})^{ij}$. Accordingly, the matrix element $\langle 0|\chi^\dagger \mathcal{K}_{^1D_2}\psi(\mu_f)|\eta_{Q2}\rangle$ is consistently associated with the same $^1D_2$ current, and the velocity-RG evolution of its Wilson coefficient $c^0_{2,2}(v)$ is governed by the corresponding anomalous dimension~\cite{Luke:1999kz}. In general, we have
\begin{eqnarray}
 v\frac{d}{d v}\ln c^S_{J,L}( v)\equiv \gamma_{J,L,S}( v),
\end{eqnarray}
where the anomalous dimension $\gamma_{J,L,S}$ has a relation of the current renormalization constant $Z_c$, i.e.,
$\gamma_{J,L,S}( v)=4\delta z^{\rm NLL}_{c,J,L,S}( v)$ where $Z_c=1+{\delta z^{\rm NLL}_{c,J,L,S}/\epsilon}+\cdots$ in $d=3-2\epsilon$ dimensions. The current renormalization constant $Z_c$ has been calculated up to next-to-leading  logarithmic (NLL) level~\cite{Hoang:2006ty}. And for the $^1D_2$ state relevant to $\eta_{Q2}\to\gamma\gamma$, one has $L=2$, $S=0$, and $J=2$. Substituting these quantum numbers into the general NLL anomalous dimension and simplifying the corresponding color and dynamical factors, it is found that
\begin{eqnarray}
\gamma_{^1D_2} = \frac{1}{5}\bigg\{ && -\frac{1}{16 \pi^{2}} \mathcal{V}_{c}^{(s)}( v)\left(\frac{\mathcal{V}_{c}^{(s)}( v)}{4}+\mathcal{V}_{r}^{(s)}( v)\right) \nonumber\\
&&+\alpha_{s}(m_Q  v)^{2}\bigg(3 \mathcal{V}_{k 1}^{(s)}( v)+2 \mathcal{V}_{k 2}^{(s)}( v)\bigg) \nonumber\\
&&+ \frac{1}{2}\alpha_{s}^{2}(m_Q  v)\left(\frac{C_{F}^{2}}{2}-C_{A} C_{F}\right)\bigg\}, 
\end{eqnarray}
where $\gamma_{^1D_2}=\gamma_{2,2,0}(v)$ is the short notation. Here $C_A=N_c$ and $C_F=(N_c^2-1)/(2N_c)$ for $SU_{C}(N_c)$ color group. Here $\mathcal{V}_{c}^{(s)}$ is the color-singlet spin-independent Coulomb potential Wilson coefficients with the operator proportional to $1/{\bf k}^2$, where ${\bf k}={\bf p}^{\prime}-{\bf  p}$ is the momentum transfer between the quark and antiquark. The coefficient $\mathcal{V}_{r}^{(s)}$ corresponds to relativistic corrections from operators proportional to $({\bf p}^{2}+{\bf p}^{\prime 2})/(2m^2{\bf k}^{2})$~\cite{Manohar:1999xd, Manohar:2000hj}
\begin{eqnarray}
\mathcal{V}_{c}^{(s)}( v) &=&-4\pi C_F\alpha_s(m_Q v),\\
\mathcal{V}_{r}^{(s)}( v) &=& -4 \pi C_F \alpha_s(m_Q v) \left[ 1 - \frac{8 C_A}{3 \beta_0} \ln(w) \right],
\end{eqnarray}
where $w=\alpha_s(m_Q v^2)/\alpha_s(m_Q v)$, and the one-loop QCD $\beta$-function coefficient $\beta_0=11-\frac{2}{3}(n_L+n_H)$. The coefficients $\mathcal{V}_{k1}^{(s)}$ and $\mathcal{V}_{k2}^{(s)}$ represent the Wilson coefficients of the color-singlet sum operators $\mathcal{O}^{(1)}_{k1}$ and $\mathcal{O}^{(T)}_{k2}$, respectively~\cite{Hoang:2002yy,Hoang:2003ns}. More explicitly,
\begin{eqnarray}
\mathcal{V}_{k1}^{(s)}( v) &=& \frac{8 C_A C_1}{3 \beta_0} \ln(w),\\
\mathcal{V}_{k2}^{(s)}( v) &=& \frac{2 C_A C_F (C_A + C_d)}{3 \beta_0} \ln(w),
\end{eqnarray}
where $C_1=C_FC_A/2-C^2_F$ and $C_d=8C_F-3C_A$. At the matching point $ v=1$, the anomalous dimension for the $^1D_2$ current reduces to:
\begin{eqnarray}
	\gamma_{^1D_2}=-\left(\frac{C_AC_F}{10}+\frac{C^2_F}{5}\right) \alpha^2_s(m_Q).
\end{eqnarray}

After converting the velocity parameter $ v$ into the factorization scale $\mu_f\sim m_Q v$, the LDME for the $^{1}D_2$ state satisfies the following evolution equation between some initial scale $\mu_{f_0}$ and the scale $\mu_f$:
\begin{eqnarray}
& & \ln\frac{\langle 0|\chi^\dag{\cal K}_{^1D_2}\psi(\mu_f)|\eta_{Q2}\rangle}{\langle 0|\chi^\dag{\cal K}_{^1D_2}\psi(\mu_{f0})|\eta_{Q2}\rangle} \nonumber\\
&=& \frac{(2C^{2}_{F}+C_F C_A)}{20} \ln\left(\frac{\mu^{2}_f} {\mu^{2}_{f_{0}}}\right)\alpha^{2}_{s}(\mu_f),
\label{eq27}
\end{eqnarray}
The total decay width \eqref{eq4} can then be expressed as
\begin{eqnarray}
\Gamma_{\eta_{Q2}\to\gamma\gamma} &=& \frac{1}{5}\frac{1}{8\pi} \frac{ |{\cal C}^{\eta_{Q2}}_{1,1}(\mu_r,\mu_f)|^2}{m_Q^5} |\langle 0|\chi^\dag{\cal K}_{^1D_2}\psi(\mu_{f_0})|\eta_{Q2}\rangle|^2 \nonumber\\
&\times& \exp\left[\frac{C_F(2C_F + C_A)}{20} \ln\left(\frac{\mu^{2}_f}{\mu^{2}_{f_{0}}}\right)\alpha^{2}_s(\mu_f) \right].
\label{eq13}
\end{eqnarray}
The universal color-singlet $\eta_{Q2}$ LDME at the initial scale $\mu_{f_0}$ can be fixed by experimental data or be related to the second derivative of the radial wavefunction at the origin via the following relation: 
\begin{eqnarray}
\langle 0|\chi^\dag{\cal K}_{^1D_2}\psi(\mu_{f_0})|\eta_{Q2}\rangle = \sqrt{\frac{5N_c}{8\pi}}{\cal R}''_D(0).
\end{eqnarray}

It is noted that by further using the scale-displacement relation for the running coupling $\alpha_{s}(\mu_{f})$ in terms of $\alpha_{s}(\mu_{r})$, the $\mu_{f}$-dependence in the short-distance coefficient $\mathcal{C}^{\eta_{Q2}}_{1,1}$ can be canceled by the corresponding $\mu_{f}$-dependence in the LDME. Since Eq.~\eqref{eq27} is the solution to the RGE for the LDME, this cancellation is equivalent to using the LDME RGE to determine its effective value. This confirms the previous observation given in Ref.~\cite{Ran:2026uhd} that the factorization-scale dependence of the fixed-order series for the total decay width can be exactly canceled by the proper use of the PMC to deal with the scale-running behavior of the LDME.

As the second step, one can further eliminate the renormalization scale dependence by employing the RGE for $\alpha_s$. For the sake of self-consistency, we outline the key procedure in what follows.

The total decay width \eqref{eq13} expanded up to $\mathcal{O}(\alpha_s^2)$ can be reexpressed as
\begin{eqnarray}
\Gamma_{\eta_{Q2}\to\gamma\gamma} = r_0\left[1 + r_1\alpha_s(\mu_r) + r_2(\mu_r)\alpha_s^2(\mu_r)\right], \label{convseries}
\end{eqnarray}
where $r_0 = \alpha^2 e_Q^4 N_c\big|{\cal R}''_D(0)\big|^2 /6 m_Q^6$, and the electromagnetic coupling constant $\alpha=1/132$ for $\eta_{c2}$ and $\alpha=1/131$ for $\eta_{b2}$. The coefficients $r_i$ can be decomposed into conformal ($r_{i,0}$) and nonconformal ($r_{i,j\neq0}$) parts using the degeneracy relations~\cite{Bi:2015wea}:
\begin{eqnarray}
r_1 &=& r_{1,0}, \\
r_2 &=& r_{2,0} + r_{2,1}\beta_0.
\end{eqnarray}
Here $n_H=1$, $n_L=3$ for $\eta_{c2}$ and $n_L=4$ for $\eta_{b2}$, respectively. Using the analytic NNLO expression given in Ref.\cite{Yang:2020pyh}, we obtain
\begin{eqnarray}
r_{1,0} &=& -0.9281C_F, \\
r_{2,0} &=& \left[-0.8856 C_A - 0.9692 C_F + 0.0433 T_F\,\sum_{i}^{n_L}\frac{e^{2}_{i}}{e^{2}_{Q}}  \right. \nonumber\\
&& \left. \quad +4.8362 T_F - 0.2898 n_H T_F  \right. \nonumber\\
&& \left. \quad +\big(0.1 C_A + 0.2 C_F\big) \ln \frac{m_Q^2}{\mu_{f_0}^2}  \right] C_F, \nonumber\\
r_{2,1} &=& -\left(0.4397 T_F + 0.0739\ln \frac{\mu_r^2}{m_Q^2} \right)\,C_F,
\end{eqnarray}
where $T_{F}=1/2$ and $N_{c}=3$ is the $\mathrm{SU}(3)_{c}$ color number. $e_{Q}$ ($Q=c,b$) represents the heavy-quark electric charge.

The series~\eqref{convseries} explicitly depends on the renormalization scale $\mu_r$, originating from mismatches between $\alpha_s$ and its corresponding expansion coefficients. In previous works, the associated theoretical uncertainties were estimated by varying the renormalization scale within the range $\mu_{r}\in[1~\mathrm{GeV},2m_{Q}]$. However, at $\mu_r=1~\mathrm{GeV}$, the perturbative suppression provided by $\alpha_s(1~\mathrm{GeV})\sim 0.48$~\cite{ParticleDataGroup:2024cfk} is insufficient to suppress the rapidly growing higher-order expansion coefficients, particularly the divergent renormalon contributions emerging at high perturbative orders~\cite{Beneke:1994qe, Neubert:1994vb}. The sizable negative NNLO corrections consequently produce an unphysical negative net result for the initial series~\eqref{convseries}~\footnote{This unphysical behavior demonstrates that obtaining reliable and precise pQCD predictions demands a rigorous scale-setting procedure, which is equally important as computing higher-order loop corrections.}. To yield physically sensible conventional predictions and exclude unphysical negative decay widths, the renormalization scale $\mu_r$ must be assigned a sufficiently large value. Accordingly, we adopt $\mu_r=2m_Q$ as the central scale within the conventional scale-setting scheme, and quantify the renormalization-scale uncertainty by varying $\mu_r$ over the interval $[m_Q,3m_Q]$.

On the other hand, by applying the standard procedures of PMC, all RGE-related nonconformal $\{\beta_i\}$ terms are absorbed into the effective coupling $\alpha_s(Q_\ast)$ for the $\eta_{Q2}\to\gamma\gamma$ process. The resulting series reads
\begin{align}
\Gamma_{\eta_{Q2}\to\gamma\gamma} &= r_0 + r_{1,0}\alpha_s(Q_\ast) + r_{2,0}\alpha_s^2(Q_\ast), \label{PMC-Gamma}
\end{align}
which is conformal and both scheme and scale invariant, thereby eliminating the above unphysical artifacts induced by adopting small yet formally perturbative renormalization scales. This invariance arises because the RGE implementation properly tunes the effective value of $\alpha_s$, such that its magnitude consistently matches the expansion coefficients of the conformal series and yields scheme-independent predictions at any fixed perturbative order. Based on the available NNLO series, the PMC scale $Q_\ast$, which corresponds to the effective momentum transfer of the decay process, is determined at the leading-logarithmic (LL) order:
\begin{eqnarray}
\ln\frac{Q_\ast^2}{m_Q^2}
&=&
-\frac{\hat{r}_{2,1}}{\hat{r}_{1,0}},
\label{eq:pmc_scale}
\end{eqnarray}
where $\hat{r}_{i,j}=r_{i,j}\big|_{\mu_r=m_Q}$.

\section{Numerical results and discussion}

For the numerical calculations performed below, we adopt the heavy-quark pole masses $m_{c}=1.67\pm0.07~\mathrm{GeV}$ and $m_{b}=4.78\pm0.06~\mathrm{GeV}$~\cite{ParticleDataGroup:2024cfk}. The initial factorization scale $\mu_{f_0}$ is conventionally of $\mathcal{O}(1~\mathrm{GeV})$, and we fix $\mu_{f_0}=1~\mathrm{GeV}$ for definiteness. The quantity $|\mathcal{R}''_{D}(0)|^{2}$ can be evaluated within an appropriate potential model. Adopting numerical results from the Cornell potential model given in Refs.\cite{Eichten:1995ch,Eichten:2019hbb},
\begin{align*}
\big|\langle 0|\chi^{\dagger} \mathcal{K}_{^{1}D_{2}} \psi(\mu_{f_0})|\eta_{c2}\rangle\big|^{2} &\approx 0.0196~\mathrm{GeV}^{7},\\
\big|\langle 0|\chi^{\dagger} \mathcal{K}_{^{1}D_{2}} \psi(\mu_{f_0})|\eta_{b2}\rangle\big|^{2} &\approx 0.5010~\mathrm{GeV}^{7},
\end{align*}
and employing the evolution equation~\eqref{eq27}, we extract the LDMEs at arbitrary factorization scales. As illustrative numerical examples, we obtain
\begin{align*}
\big|\langle 0|\chi^{\dagger} \mathcal{K}_{^{1}D_{2}} \psi(\mu_{f}=m_{c})|\eta_{c2}\rangle\big|^{2} &\approx 0.0215~\mathrm{GeV}^{7},\\
\big|\langle 0|\chi^{\dagger} \mathcal{K}_{^{1}D_{2}} \psi(\mu_{f}=m_{b})|\eta_{b2}\rangle\big|^{2} &\approx 0.5593~\mathrm{GeV}^{7},
\end{align*}
for the $\eta_{c2}$ and $\eta_{b2}$ states, respectively. 

\subsection{Total decay width of $\eta_{c2}$}

For the $\eta_{c2}\to\gamma\gamma$ decay process with $m_Q = m_c$, the PMC scale determined from Eq.~\eqref{eq:pmc_scale} reads
\begin{displaymath}
Q_{\ast} = 1.483~\mathrm{GeV}.
\end{displaymath} 

\begin{figure}[htb]
\centering
\includegraphics[width=0.48\textwidth]{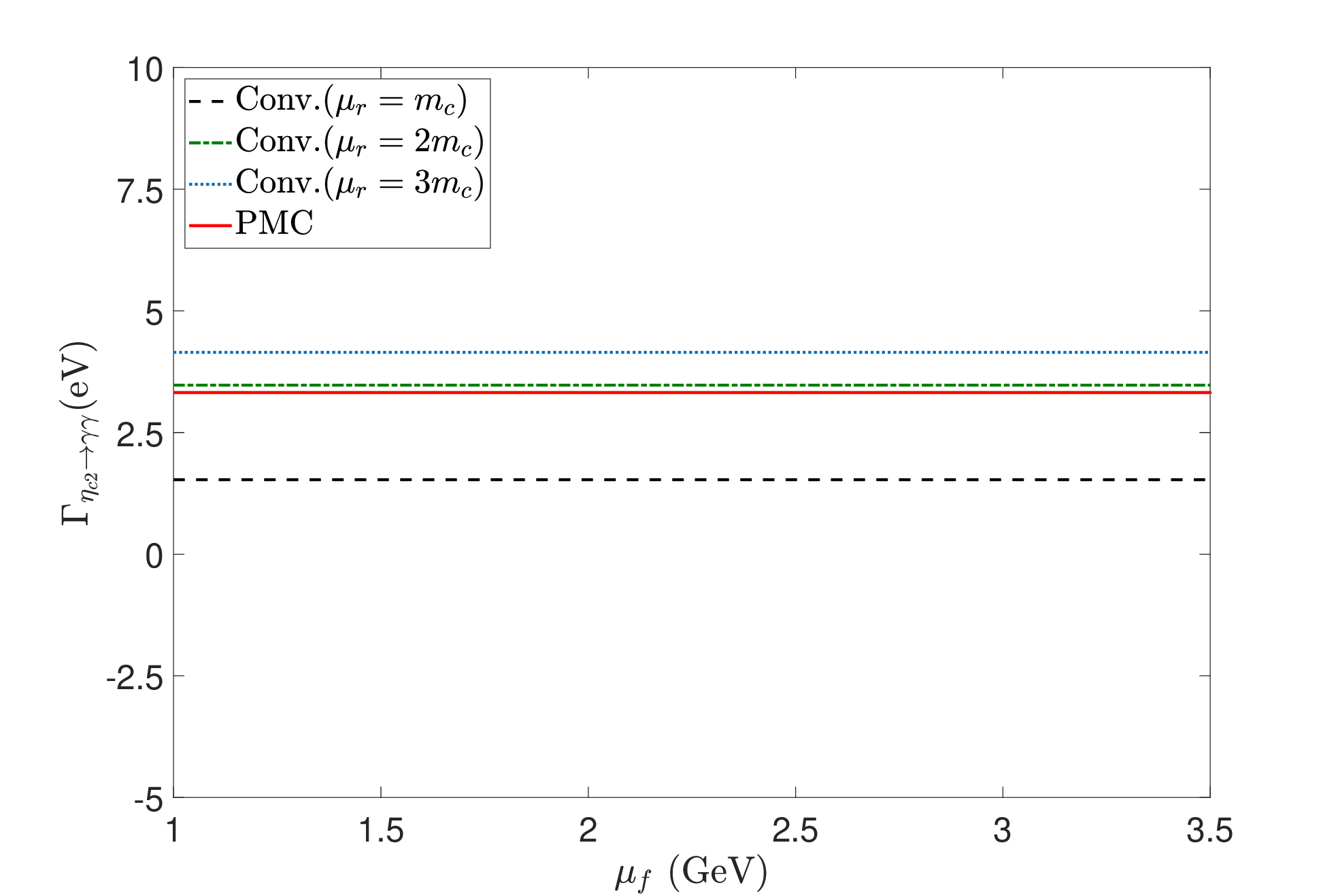}
\caption{The factorization-scale $\mu_f$ dependence of the total decay width of $\eta_{c2}$, including the scale evolution of the $D$-wave LDME, obtained within the conventional (Conv.) and PMC scale-setting methods. }
\label{figo}
\end{figure}

Fig.~\ref{figo} shows the factorization-scale dependence of the $\eta_{c2}$ decay width after incorporating the evolution effects of the relevant LDME $\langle 0|\chi^{\dagger}{\cal K}_{^{1}\!D_{2}}\psi(\mu_{f})|\eta_{c2}\rangle$. Compared with previous results in the literature, we demonstrate that the factorization-scale dependence of the $\eta_{c2}$ decay width can be completely eliminated once the scale evolution of the $D$-wave LDME is properly included. Such cancellation of the $\mu_f$ dependence between the perturbative hard-scattering part and the nonperturbative hadronic matrix element yields much more stable and reliable theoretical predictions for the $\eta_{c2}$ decay width.

\begin{figure}[htb]
\centering
\includegraphics[width=0.48\textwidth]{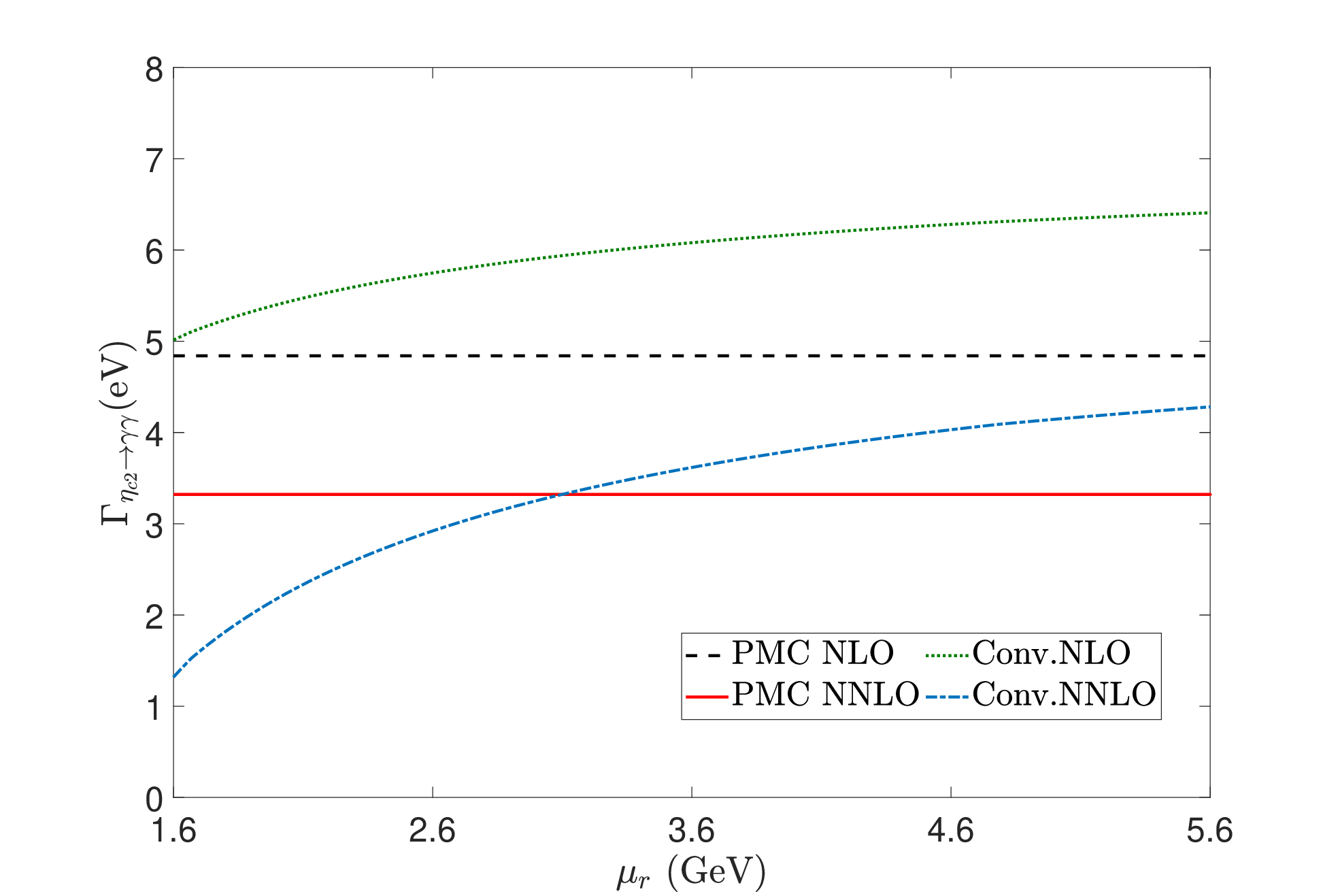}
\caption{The total decay width $\Gamma_{\eta_{c2}\to\gamma\gamma}$ as a function of the renormalization scale $\mu_r$. Results from the conventional (Conv.) and PMC scale-setting methods are shown at NLO and NNLO accuracy, respectively. }
\label{fig1}
\end{figure}

As discussed above, scale evolution of the LDME eliminates the factorization-scale dependence, leaving the renormalization scale as the dominant source of theoretical uncertainty for fixed-order predictions within the conventional scale-setting method (Conv.). To quantify the renormalization-scale uncertainty, we display the $\mu_r$ dependence of the total decay width $\Gamma_{\eta_{c2}\to\gamma\gamma}$ before and after implementing the PMC method in Fig.~\ref{fig1}. The dotted and dot-dashed lines denote conventional predictions at NLO and NNLO accuracy, while the dashed and solid lines correspond to PMC predictions at NLO and NNLO accuracy, respectively. The conventional result $\Gamma_{\eta_{c2}\to\gamma\gamma}\big|_{\rm Conv.}$ exhibits strong sensitivity to the renormalization scale $\mu_r$. Varying $\mu_r\in[m_c,3m_c]$, the total decay width varies within the range $[1.530,4.147]~\mathrm{eV}$. For the conventional method, the individual LO, NLO, and NNLO contributions (in units of eV) at different central renormalization scales are presented as
\begin{align}
\label{eq16}
{\rm Conv.}\big|_{\mu_r=m_c} &= \{8.597,\,-3.492,\,-3.575\},\\
\label{eq17}
{\rm Conv.}\big|_{\mu_r=2m_c} &= \{8.597,\,-2.586,\,-2.538\},\\
\label{eq18}
{\rm Conv.}\big|_{\mu_r=3m_c} &= \{8.597,\,-2.258,\,-2.192\},
\end{align}
where the three numerical entries correspond to the contributions of the LO-terms, the NLO-terms, and the NNLO-terms, respectively. The conventional results show sizable renormalization-scale uncertainties, amounting to $\left(^{-12.7\%}_{+35.0\%}\right)$ for the NLO-terms and $\left(^{-13.6\%}_{+40.9\%}\right)$ for the NNLO-terms. Furthermore, there is no cancellation between contributions of different perturbative orders, which causes the overall renormalization-scale uncertainty to grow for the NNLO series.

In contrast, the $\mu_r$ dependence is effectively eliminated for PMC predictions of the decay process $\eta_{c2}\to\gamma\gamma$. The pQCD corrections to the total decay width~\eqref{PMC-Gamma} obtained via the PMC method at each perturbative order read
\begin{align}
\label{eq19}
{\rm PMC} &= \{8.597,\,-3.757,\,-1.518\},
\end{align}
and this set of values remains invariant under any choice of $\mu_r$ order by order. This demonstrates that PMC yields more robust theoretical predictions. Furthermore, since PMC resums the divergent renormalon contributions, the resulting perturbative series exhibits improved convergence of its expansion coefficients.

\begin{table}[htb]
\begin{center}
\begin{tabular}{c c c c c c}
\hline
& ~$\eta_{c2}$~ & ~$ k_{1} $~ & ~$ k_{2} $\\
\hline
& $\rm Conv.$ & $-30.1\%^{+3.8\%}_{-10.5\%}$ & $-29.5\%^{+4.0\%}_{-12.1\%}$  \\
& $\rm PMC$ & $-43.7\%$ & $-17.7\%$ \\
\hline
\end{tabular}
\caption{The ratio $k_{1,2}$ obtained using the conventional method (with $\mu_{r}=2m_{c}$) and the PMC method. The uncertainty in the conventional result is estimated by varying the renormalization scale in the range $\mu_{r}\in[m_{c},\,3m_{c}]$.}
\label{tab1}
\end{center}
\end{table}

To characterize the perturbative convergence for both the conventional and the PMC methods, we define the following ratio $k_{i}$,
\begin{equation}
k_{i} = \frac{\mathrm{N^{i}LO}}{\mathrm{LO}} \times 100\%,
\end{equation}
where $i=1,2$ corresponds to the NLO and NNLO corrections, respectively. The corresponding predictions are summarized in Table~\ref{tab1}. It is evident that the conventional perturbative series exhibits poor convergence. For example, when the renormalization scale is chosen to be $\mu_{r}=2m_{c}$, the NNLO contribution becomes comparable in magnitude to the NLO contribution, e.g., 
\begin{align}
k_{1}|_{\mathrm{Conv.}} &= -30.1\%^{+3.8\%}_{-10.5\%}, \\
k_{2}|_{\mathrm{Conv.}} &= -29.5\%^{+4.0\%}_{-12.1\%},
\end{align}
where the uncertainties are estimated by varying the renormalization scale in the range $\mu_{r}\in[m_{c},\,3m_{c}]$. In contrast, the PMC prediction exhibits an improved scale-invariant perturbative convergence, with
\begin{equation}
k_{1}|_{\mathrm{PMC}} = -43.7\%, \qquad k_{2}|_{\mathrm{PMC}} = -17.7\%.
\end{equation}
The NNLO correction is suppressed relative to the NLO term. Consequently, the scale-invariant PMC series can be taken as a more accurate reflection of the intrinsic properties of perturbative QCD.

Ultimately, the total decay width of $\eta_{c2} \to \gamma\gamma$ before and after applying the PMC method is:
\begin{align}
\Gamma_{\eta_{c2}\to\gamma\gamma}|_{\mathrm{Conv.}} &= 3.473^{+0.674}_{-1.943}~\mathrm{eV}, \\
\Gamma_{\eta_{c2}\to\gamma\gamma}|_{\mathrm{PMC}} &= 3.322~\mathrm{eV},
\end{align}
where the central value of conventional result is fixed at $\mu_{r}=2m_{c}$, and the uncertainties are for $\mu_{r} \in [m_{c},\,3m_{c}]$.

\begin{figure}[htb]
\centering
\includegraphics[width=0.48\textwidth]{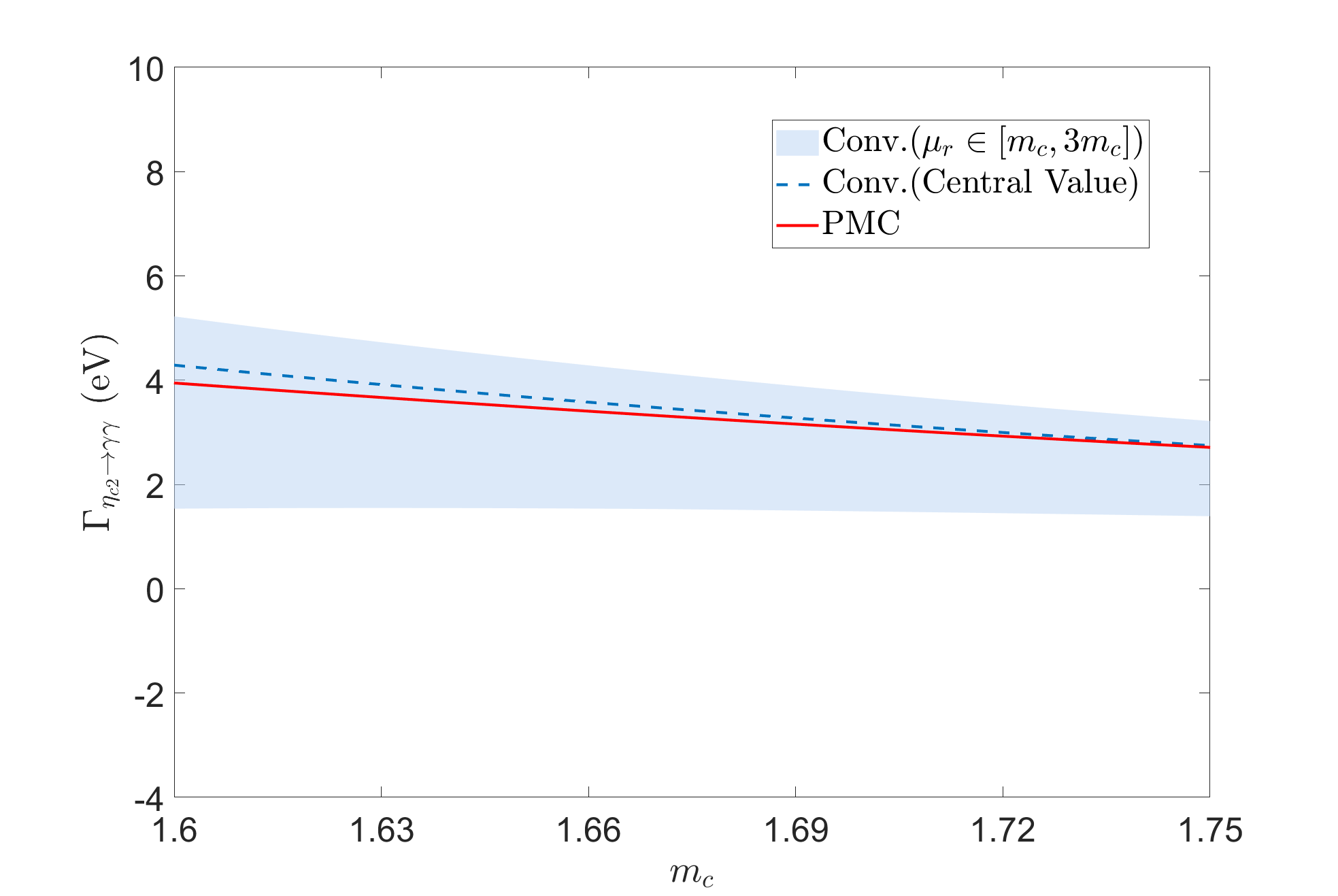}
\caption{Variation of the total decay width $\Gamma_{\eta_{c2}\to\gamma\gamma}$ with the charm quark pole mass $m_c = 1.67 \pm 0.07~\mathrm{GeV}$ \cite{ParticleDataGroup:2024cfk}. }
\label{figmass}
\end{figure}

As a final remark, we evaluate the uncertainty originating from the charm-quark pole mass by varying $m_c$ within the PDG range $m_c = 1.67 \pm 0.07~\mathrm{GeV}$~\cite{ParticleDataGroup:2024cfk}. The corresponding numerical results are presented in Fig.~\ref{figmass}. Quantitatively, we obtain
\begin{equation}
\Gamma_{\eta_{c2}\to\gamma\gamma}|^{\Delta m_c}_{\mathrm{PMC}} = 3.322^{+0.656}_{-0.555}~\mathrm{eV}
\end{equation}
for the PMC series. As a comparison, the mass-induced uncertainties for the pQCD series \eqref{eq13} under conventional scale-setting method are 
\begin{align}
\Gamma_{\eta_{c2}\to\gamma\gamma}|^{\Delta m_c}_{\mathrm{Conv.}}(\mu_r = m_c) &= 1.530^{+0.050}_{-0.118}~\mathrm{eV},\\
\Gamma_{\eta_{c2}\to\gamma\gamma}|^{\Delta m_c}_{\mathrm{Conv.}}(\mu_r = 2m_c) &= 3.473^{+0.815}_{-0.646}~\mathrm{eV},\\
\Gamma_{\eta_{c2}\to\gamma\gamma}|^{\Delta m_c}_{\mathrm{Conv.}}(\mu_r = 3m_c) &= 4.147^{+1.077}_{-0.828}~\mathrm{eV}.
\end{align}
These results demonstrate that the uncertainty in the charm-quark pole mass remains a dominant source of theoretical error for the $\eta_{c2}$ total decay width. In contrast to the conventional scale-setting method, for which both the central value and theoretical uncertainty strongly depend on the renormalization scale $\mu_r$, the PMC scale-setting method yields robust mass-induced uncertainty estimates. Specifically, the mass uncertainty originating from $\Delta m_c = \pm 0.07~\mathrm{GeV}$ leads to a scale-invariant uncertainty of $\left(^{+20\%}_{-17\%}\right)$ for the PMC series.

\subsection{Total decay width of $\eta_{b2}$}

\begin{figure}[htb]
\centering
\includegraphics[width=0.48\textwidth]{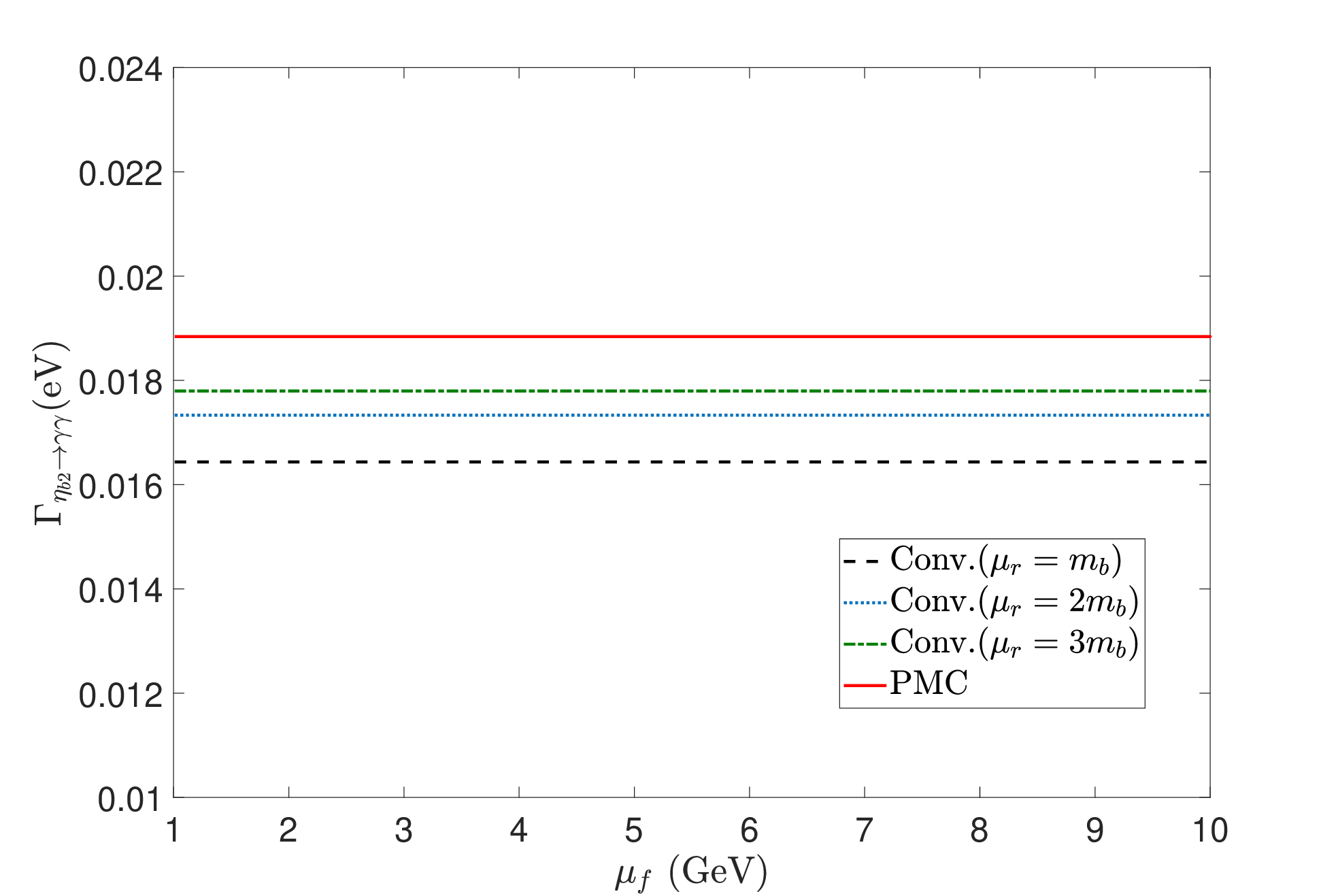}
\caption{Factorization scale $\mu_f$ dependence of $\Gamma_{\eta_{b2}\to\gamma\gamma}$ under the conventional (Conv.) and PMC scale-setting methods.}
\label{figbf}
\end{figure}

\begin{figure}[htb]
\centering
\includegraphics[width=0.48\textwidth]{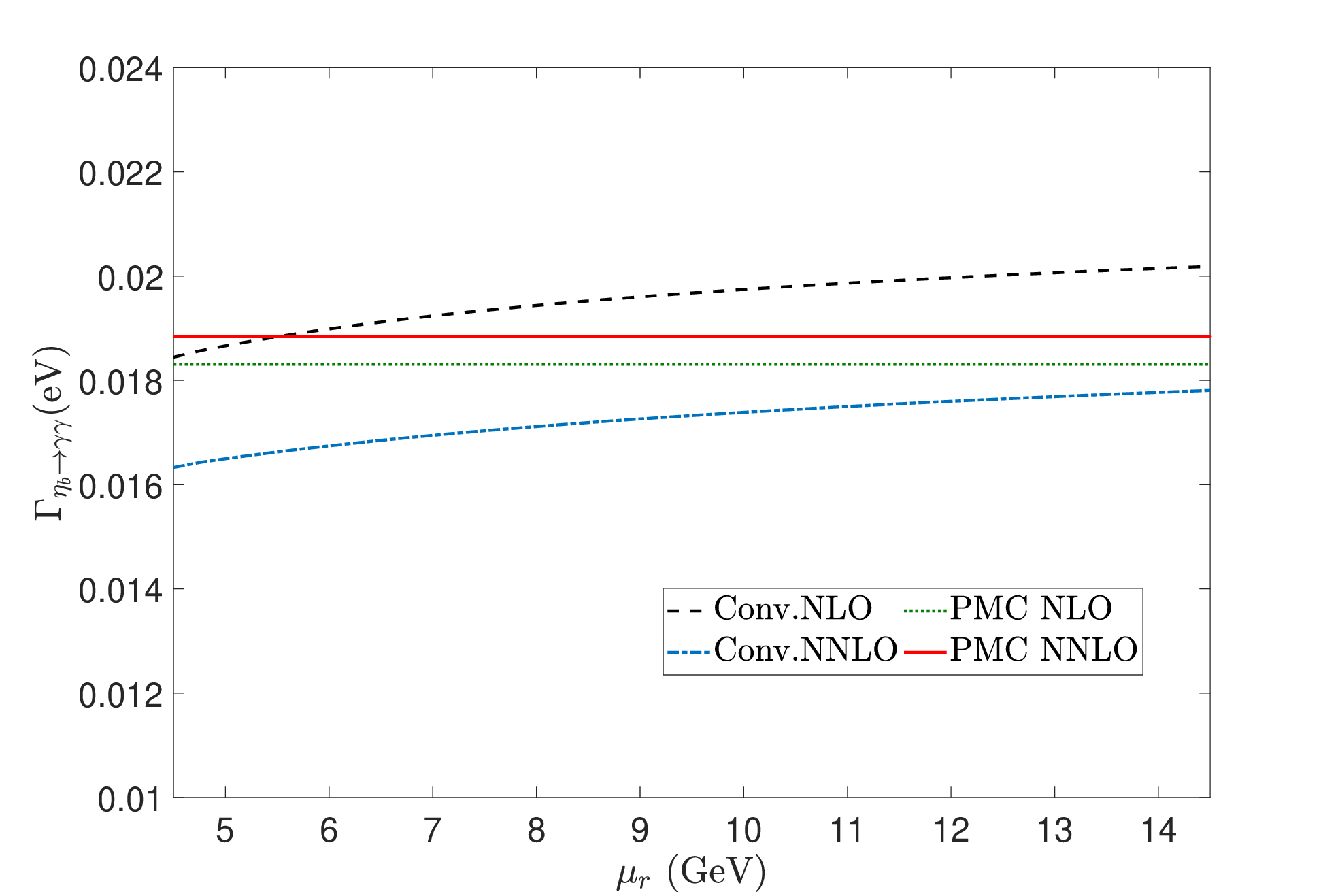}
\caption{Total decay width of $\eta_{b2}\to\gamma\gamma$ up to NLO and NNLO as a function of the renormalization scale $\mu_r$ using the conventional (Conv.) and PMC scale-setting methods.}
\label{figff}
\end{figure}

\begin{table}[htb]
\centering
\renewcommand{\arraystretch}{1.2} 
\begin{tabular}{c c c c c} 
\toprule
\multicolumn{1}{c}{$\Gamma_{\eta_{b2}}~(\mathrm{eV})$} & LO & NLO & NNLO & Total \\
\midrule
Conv. & $0.0253$ & $-0.0056^{+0.0005}_{-0.0011}$ & $-0.0023^{-0.0001}_{+0.0002}$ & $0.0174^{+0.0004}_{-0.0009}$ \\
PMC   & $0.0253$ & $-0.0070$ & $0.0005$ & $0.0188$  \\
\bottomrule
\end{tabular}
\caption{The N$^i$LO-term contributions to the decay width $\Gamma_{\eta_{b2}\to\gamma\gamma}$ under the PMC and conventional scale-setting methods. The $\mu_r$-uncertainty of the conventional results is determined by varying the scale within the range $\mu_r\in[m_b,\,3m_b]$.}
\label{tab3}
\end{table}

The theoretical framework presented above can be straightforwardly applied to describe bottomonium $\eta_{b2}$ decays into two photons, and these processes may be measured in future high-precision experiments. The factorization scale dependence of the total decay width $\Gamma_{\eta_{b2}\to\gamma\gamma}$ is investigated. As demonstrated in Fig.~\ref{figbf}, incorporating the LDME evolution eliminates the factorization scale dependence of the $\eta_{b2}$ total decay width at the considered accuracy, which is the same as the $\eta_{c2}$ case. With the factorization-scale ambiguity removed, we now turn to the uncertainty of the renormalization scale. 

For the $\eta_{b2}\to\gamma\gamma$, the PMC scale is determined to be $Q_\ast=4.246$ GeV. Figure~\ref{figff} shows the contributions to the decay width $\Gamma_{\eta_{b2}\to\gamma\gamma}$ up to NLO and NNLO levels under the conventional and PMC scale-setting methods. Table~\ref{tab3} demonstrates the numerical comparison of the conventional and PMC scale-setting methods for the total decay width $\Gamma_{\eta_{b2}\to\gamma\gamma}$ order by order. Due to the relatively small coupling $\alpha_s(m_b) \sim 0.2$, the perturbative expansion for the bottomonium system is well behaved and exhibits good convergence. However, the conventional prediction still retains a sizable dependence on the renormalization scale. Specifically, after varying the renormalization scale within $\mu_r \in [m_b,\,3m_b]$, the uncertainty of the total decay width under the conventional scale-setting method is $\left(^{+2.3\%}_{-5.2\%}\right)$. In contrast, the PMC series is scale-invariant and demonstrates better convergence behavior. The numerical result for the bottomonium decay width is $\Gamma_{\eta_{b2}\to\gamma\gamma}|_{\mathrm{PMC}} = 0.0188~\mathrm{eV}$.

Similarly, after considering the bottom quark pole mass uncertainty $m_b = 4.78 \pm 0.06~\mathrm{GeV}$ \cite{ParticleDataGroup:2024cfk}, we have 
\begin{align}
\Gamma_{\eta_{b2}\to\gamma\gamma}|^{\Delta m_b}_{\mathrm{Conv.}}(\mu_r = m_b) &= 0.0165^{+0.0012}_{-0.0011}~\mathrm{eV},\\
\Gamma_{\eta_{b2}\to\gamma\gamma}|^{\Delta m_b}_{\mathrm{Conv.}}(\mu_r = 2m_b) &= 0.0174^{+0.0013}_{-0.0012}~\mathrm{eV},\\
\Gamma_{\eta_{b2}\to\gamma\gamma}|^{\Delta m_b}_{\mathrm{Conv.}}(\mu_r = 3m_b) &= 0.0178^{+0.0014}_{-0.0012}~\mathrm{eV}
\end{align}
for the conventional scale-setting method. By contrast, the PMC prediction is
\begin{equation}
\Gamma_{\eta_{b2}\to\gamma\gamma}|^{\Delta m_b}_{\mathrm{PMC}} = 0.0188^{+0.0014}_{-0.0013}~\mathrm{eV}.
\end{equation}

\subsection{An estimation of $\mathrm{N}^3\mathrm{LO}$ contribution using the Pad\'e approximation approach}

It is useful to estimate contributions from unknown higher-order (UHO) terms, which constitute an intrinsic, unavoidable theoretical uncertainty inherent to perturbation theory. The scale-invariant conformal coefficients derived via the PMC enable us to construct more reliable estimates for UHO corrections. In what follows, to further investigate the convergence properties of the perturbative series, we employ the Pad\'e approximation approach (PAA)~\cite{Basdevant:1972fe, Samuel:1992qg, Samuel:1995jc} -- a resummation technique based on fractional generating functions -- to quantify the uncomputed $\mathrm{N}^3\mathrm{LO}$ contribution to the total decay width of the process $\eta_{c2} \to \gamma\gamma$. 

Under the PAA, the uncalculated N$^3$LO contribution is encoded in a $[N/M]$-type rational function. To match with the present NNLO analysis, we adopt the $[0/1]$-type PAA to systematically estimate the higher-order contributions~\cite{Du:2018dma}, which is consistent with the PMC procedures and yields the perturbative coefficients for the missing N$^3$LO-term as follows:
\begin{equation}
r_3^{\rm PAA} = \frac{r^{2}_{2}}{r_{1}},
\end{equation}
for the conventional scale-setting method and 
\begin{equation}
r_{3,0}^{\rm PAA} = \frac{r^{2}_{2,0}}{r_{1,0}},
\end{equation}
for the PMC. Numerically, the predicted N$^3$LO contributions to $\Gamma_{\eta_{c2}\to\gamma\gamma}$ are
\begin{equation}
\Gamma_{\eta_{c2}\to\gamma\gamma}|^{\rm N^3LO}_{\mathrm{Conv.+PAA}} = -2.490^{+0.362}_{-1.169}~\mathrm{eV},
\end{equation}
for the conventional scale-setting method, and the PMC result is
\begin{equation}
\Gamma_{\eta_{c2}\to\gamma\gamma}|^{\rm N^3LO}_{\mathrm{PMC+PAA}} = -0.614~\mathrm{eV}.
\end{equation}
The central value of the conventional result is obtained at $\mu_r = 2m_c$, where the error comes from the renormalization scale uncertainty ($\mu_r \in [m_c,\,3m_c]$). The PMC renormalization scale-invariant result removes the renormalization scale uncertainty and significantly enhances the accuracy of the N$^3$LO-term prediction.

Meanwhile, we also investigate the case of the bottom quark. Since the strong coupling is smaller in the bottom-quark case, the numerical results show that the N$^3$LO contribution is sufficiently suppressed. For the conventional scale-setting method, we have
\begin{equation}
\Gamma_{\eta_{b2}\to\gamma\gamma}|^{\rm N^3LO}_{\mathrm{Conv.+PAA}} = -0.0010^{+0.0003}_{-0.0001}~\mathrm{eV},
\end{equation}
and the PMC result is
\begin{equation}
\Gamma_{\eta_{b2}\to\gamma\gamma}|^{\rm N^3LO}_{\mathrm{PMC+PAA}} = -0.0001~\mathrm{eV}.
\end{equation}

\subsection{Branching ratios}

By incorporating the combined uncertainties from the renormalization scale and the UHO terms with the quark mass effects ($\Delta m_c = \pm 0.07\;\mathrm{GeV}$ for $\eta_{c2}$ decays and $\Delta m_b = \pm 0.06\;\mathrm{GeV}$ for $\eta_{b2}$ decays), we obtain the total theoretical predictions for the $\eta_{Q2}$ decay width. Under the conventional scale-setting method, the results are
\begin{align}
\Gamma_{\eta_{c2}\to\gamma\gamma}\big|_{\mathrm{Conv.}} &= 3.473^{+2.705}_{-3.224}\;\mathrm{eV}, \\
\Gamma_{\eta_{b2}\to\gamma\gamma}\big|_{\mathrm{Conv.}} &= 0.0174^{+0.0017}_{-0.0018}\;\mathrm{eV},
\end{align}
while the PMC predictions read
\begin{align}
\Gamma_{\eta_{c2}\to\gamma\gamma}\big|_{\mathrm{PMC}} &= 3.322^{+0.899}_{-0.828}\;\mathrm{eV}, \\
\Gamma_{\eta_{b2}\to\gamma\gamma}\big|_{\mathrm{PMC}} &= 0.0188^{+0.0014}_{-0.0013}\;\mathrm{eV}.
\end{align}
The total uncertainty of the PMC series is $\sim \left(^{+27.1\%}_{-24.9\%}\right)$ for $\eta_{c2}$ decays and $\sim \left(^{+7.4\%}_{-6.9\%}\right)$ for $\eta_{b2}$ decays, which are dominated by $\Delta m_Q$. In contrast, the conventional method yields a substantially larger uncertainty of $\sim \left(^{+77.8\%}_{-92.8\%}\right)$ for $\eta_{c2}$ decays and $\sim \left(^{+9.8\%}_{-10.3\%}\right)$ for $\eta_{b2}$ decays; these errors arise predominantly from renormalization scale dependence, with the scale window taken as $\mu_r \in [m_Q,\,3m_Q]$.

We further present predictions for the branching ratio $\mathrm{Br}(\eta_{Q2} \to \gamma\gamma)$, which is defined as
\begin{align}
\mathrm{Br}(\eta_{Q2} \to \gamma\gamma) = \frac{\Gamma_{\eta_{Q2} \to \gamma\gamma}}{\Gamma_{\mathrm{total}}(\eta_{Q2})},
\end{align}
where the total decay widths are dominated by E1 transitions and hadronic decays. For our subsequent analysis, we adopt $\Gamma_{\mathrm{total}}(\eta_{c2}) = 445.1\;\mathrm{keV}$ and $\Gamma_{\mathrm{total}}(\eta_{b2}) = 29.1\;\mathrm{keV}$ as quoted in Ref.~\cite{Yang:2020pyh}. Using these total widths, we compute the branching ratios within the conventional scale-setting method:
\begin{align}
\mathrm{Br}(\eta_{c2}\to \gamma\gamma)\big|_{\mathrm{Conv.}} &= \bigl(7.803^{+6.077}_{-7.243}\bigr)\times10^{-6}, \\
\mathrm{Br}(\eta_{b2}\to \gamma\gamma)\big|_{\mathrm{Conv.}} &= \bigl(5.979^{+0.584}_{-0.619}\bigr)\times10^{-7}.
\end{align}
The corresponding PMC predictions read
\begin{align}
\mathrm{Br}(\eta_{c2}\to \gamma\gamma)\big|_{\mathrm{PMC}} &= \bigl(7.463^{+2.020}_{-1.860}\bigr)\times10^{-6},\\
\mathrm{Br}(\eta_{b2}\to \gamma\gamma)\big|_{\mathrm{PMC}} &= \bigl(6.460^{+0.481}_{-0.447}\bigr)\times10^{-7}.
\end{align}

\section{Summary}

In this work, we perform a detailed analysis of pQCD corrections to the $\eta_{Q2}\to \gamma\gamma$ decay width up to NNLO accuracy within the PMC framework. By carrying out the LDME evolution, the factorization-scale uncertainty is fully eliminated. Meanwhile, by systematically absorbing all RGE-dependent non-conformal $\{\beta_i\}$ contributions into the strong coupling, the PMC fixes an effective coupling $\alpha_s(Q_*)$ with the characteristic PMC scale $Q_*$ determined at LL order. This procedure removes the conventional renormalization-scheme and scale ambiguities, accelerates the convergence of the perturbative series, becomes a better basis for estimating UHO contributions, and delivers more precise predictions compared with the conventional scale-setting method. 

The PMC predictions for the total two-photon decay widths read $\Gamma_{\eta_{c2}\to\gamma\gamma}\big|_{\mathrm{PMC}} = 3.322^{+0.899}_{-0.828}\ \mathrm{eV}$ and $\Gamma_{\eta_{b2}\to\gamma\gamma}\big|_{\mathrm{PMC}} = 0.0188^{+0.0014}_{-0.0013}\ \mathrm{eV}$. Furthermore, the theoretical uncertainties of the PMC predictions stemming from UHO terms are substantially suppressed compared with those from the conventional scale-setting method. Specifically, the uncertainty bands are reduced by $(^{+27.1\%}_{-24.9\%})$ for $\eta_{c2}$ and $(^{+7.4\%}_{-6.9\%})$ for $\eta_{b2}$, which clearly demonstrates the improved theoretical precision afforded by the PMC procedure. The predicted branching ratios reside at the $10^{-6}$ level for $\eta_{c2}$ and the $10^{-7}$ level for $\eta_{b2}$, corresponding to highly suppressed two-photon partial widths; thus, direct experimental observation of this decay mode is extremely challenging. Nevertheless, this rare electromagnetic decay remains a valuable observable for high-luminosity collider facilities, particularly for searches targeting $C$-even heavy-quarkonium states whose quantum numbers allow for two-photon production and relevant radiative processes at electron–positron colliders. These scheme- and scale-independent branching ratios provide reliable theoretical benchmarks for future experimental searches for the $\eta_{Q2}$ resonance signal.

\acknowledgments

This work was supported by the Natural Science Foundation of China under Grant No.12305091, No.12575080 and No.12547101, supported by Sichuan Science and Technology Program No. 2026NSFSC0758 and No.2024NSFSC1367, by the Research Fund for the Doctoral Program of the Southwest University of Science and Technology under Contract No.24zx7117 and No.23zx7122.

\end{document}